\documentclass[12pt]{article}
\makeatletter
\@addtoreset{equation}{section}
\makeatother

\renewcommand{\title}[1]{\null\vspace{25mm}

\noindent{\Large{\bf #1}}\vspace{10mm}

}
\newcommand{\authors}[1]{\noindent{\large #1}\vspace{3mm}

}

\renewcommand{\abstract}[1]{\vspace{19mm}

\noindent{\small{\em Abstract.} #1}\vspace{2mm}
}

\usepackage{amssymb}
\usepackage{latexsym}
\usepackage{overcite}
\usepackage{amsthm}
\theoremstyle{plain}

\makeatletter \renewcommand\@biblabel[1]{#1.} \makeatother
\newtheorem{theorem}{Theorem}[section]
\newtheorem{definition}[theorem]{Definition}
\newtheorem{lemma}[theorem]{Lemma}

\newcommand{\uline}{\vrule height.06ex depth.02ex width.6em}
\newcommand{\mvee}{\vee\kern-.69em\uline}

\font\1=cmss10
\font\2=cmssdc10 scaled\magstep2

\begin{document}

\markright{Megill and Pavi\v ci\'c (2001)}

\title{Deduction, Ordering, and Operations in Quantum Logic}

\authors{Norman D.~Megill$\footnote{Boston Information Group, 30 Church St.,
Belmont MA 02478, U.~S.~A. E-mail: nm@alum.mit.edu;
Web page: http://www.metamath.org}$
and Mladen Pavi\v ci\'c$\footnote{University of Zagreb, 
AGG Fakultet, Physics Division, Ka\v ci\'ceva 26,
HR-10000 Zagreb, Croatia.
E-mail: mpavicic@faust.irb.hr; Web page: http://m3k.grad.hr/pavicic}$}
\abstract{We show that in quantum logic of closed subspaces of Hilbert
space one cannot substitute quantum operations for classical 
(standard Hilbert space) ones and treat them as primitive operations. 
We consider two possible ways of such a substitution and arrive at 
operation algebras that are not lattices what proves the claim. 
We devise algorithms and programs which write down any two-variable 
expression in an orthomodular lattice by means of classical and 
quantum operations in an identical form. Our results show that lattice 
structure and classical operations uniquely determine quantum logic
underlying Hilbert space. As a consequence of our result, recent 
proposals for a deduction theorem with quantum operations in an 
orthomodular lattice as well as a substitution of quantum operations 
for the usual standard Hilbert space ones in quantum logic prove to be 
misleading. Quantum computer quantum logic is also discussed.}



\section{Introduction}
\label{sec:intro}

Quantum computer theory recently introduced quantum logic as an algebra
of quantum bits (qubits) handled by quantum logic gates \cite{q-com-ql}
and as a quantum counterpart of the classical Boolean algebra which
is an algebra of classical bits handled by classical logic gates
in today's computers. This quantum algebra is still not well defined
but it may eventually emerge as a variety of equations defining
quantum logic of Hilbert space---so-called Hilbert lattice (algebra
of subspaces or, equivalently, operators  of Hilbert
space)---introduced half a century ago.$^{(19,34)}$
A project of reducing the two logics to each other should first of
all address those features of the Hilbert space quantum logic whose
implementation into the computer quantum logic are hardly possible
or problematic. Such features are:
\begin{itemize}
\item The Hilbert space quantum logic (Hilbert lattice) is
(ortho)isomorphic to the lattice of the closed subspaces of any
Hilbert space. It has recently been shown that
{\it infinite orthogonality} and {\it unitarity} added to the
lattice make sure that the field over which any infinite
dimensional Hilbert space is defined must be either complex,
or real or quaternionic. This does not hold for a finite dimensional
Hilbert space which allows non-standard (e.g., Keller's \cite{keller}) 
fields that can make the space nonarchimedean. The latter possibility 
should be eliminated in favor of a complex field, but this requires 
further research since the theory of finite Hilbert lattices and of 
finite complex Hilbert space is poorly developed. Quantum computer
quantum logic obviously must be finite dimensional and therefore
we need a finite Hilbert lattice theory with conditions that
eliminate exotic properties and have a plausible physical
interpretation. Besides, a finite dimensional quantum theory might
offer us a model for discrete space which goes beyond any numerical
grid approximation of quantum state equations (as given, e.g., by
Ref.~\citen{schr-simul} for quantum computer).

\item Hilbert lattice theory contains conditions of the second
order which involve universal and existential quantifiers. Whether
they can be translated into algebraic conditions and equational series,
and as such possibly handled and approximated by quantum computer,
is an open question. An algebraic approach to the Hilbert lattices,
whose goal is to substitute the aforementioned quantifier
conditions with algebraic equations has been started only 
recently.$^{(12,13,21,23)}$

\item There was a long standing question on whether a \it proper\/
\rm logic---not its above considered algebraic model which is
misleadingly called quantum \it logic\/\rm---underlying
Hilbert space can play the same
role classical logic, underlying the Boolean algebra,
played---it was believed---in any classical physical and computer 
theory.$^{(5,6,11,16,17,25,30,33)}$
However, two years ago it was
discovered \cite{mpcommp99} that there are \it proper\/
\rm logics for neither quantum nor classical theories.
It turned out that both quantum and classical \it proper\/
\rm logics have at least two models each and that their syntax
correspond to models that are neither Hilbert lattice nor
Boolean algebra, respectively. Hence,  \it proper\/
\rm logics can characterize neither quantum nor classical
computers. More generally, they can characterize neither quantum
nor classical theories---physical as well as mathematical.
Hence, all the literature on the subject from before 1999 is
outdated by the result and all newer papers that do not take
the result into account---as, e.g., Dalla Chiara and Giuntini
\cite{dalla-c-giunt}---are simply incorrect. For, obviously, one
cannot claim that a statement holds in \it proper\/ \rm
quantum logic if and only if it is true in orthomodular lattices
and at the same time that a statement holds in \it proper\/ \rm
quantum logic if and only if it is true in non-orthomodular
lattices. The same holds for \it proper\/ \rm classical
logic. Consequently, another long standing \it proper\/ \rm
quantum logic issue---whether or not there is a \it deduction
theorem\/ \rm for the logic---is put aside by the above result.
However, the deduction theorem was recently reformulated in a lattice
theoretic framework and we will consider the latter reformulation in
the present paper. This will take us to a problem of whether there
is {\it quantum\/ \rm} lattice ordering in Hilbert lattices.

\item Orthomodular lattices---which contain Hilbert lattices---use
operations taken over from Hilbert space. However, there are
also five the so-called quantum versions of them.~\cite{mpqo01}
The latter reduce to the former for compatible variables.
For almost half a century various authors have tried to find
\it proper\/ \rm quantum operations. This will also take us to a
problem of possible \it quantum\/ \rm  ordering in Hilbert
lattices. It will turn out that in a standard formulation
of Hilbert lattices with standard operations there are
\it quantum\/ \rm  orderings. But if we wanted to use only
\it quantum\/ \rm operations and only \it quantum\/ \rm  orderings,
we would arrive at algebras that are not lattices and which
therefore cannot be made (by means of additional conditions) 
orthoisomorphic to the lattice of closed subspaces of Hilbert space. 
Hence, the standard lattice operations inherited from Hilbert space 
are the only ``proper'' operations. The result greatly simplifies 
approaches to the first two points above since it shows that any 
use of ``quantum operations'' in finding new conditions and 
equations in Hilbert lattices can only be a casual matter of 
convenience. E.g., in formulations of orthoarguesian laws some of
quantum operations appear in characterizations of equations 
\cite{mpoa99} and now we know that they only make equations shorter 
to write.
\end{itemize}

Taken together, setting down the problems with quantum operations,
ordering, and deduction will narrow the gap between the two logics
and this is what we strive at in this paper. However, before
dwelling to the task, the following historical comments might be helpful.

Thirty years ago Finch \cite{finch70} noticed that, in an orthomodular
lattice, the operation $a\cap(b\cup a')$, which we denote
as $a\cap_1 b$ and call
\it quantum conjunction\/\rm, ``could be interpreted as
an operation of logical conjunction'' and that it satisfies the following
condition: $b\cap_1 a\le c\ \Leftrightarrow\ a\le b\to_1 c$
where $a\to_1 b$ denotes $(a\cap_1 b')'$ and is called
\it quantum implication\/\rm. Rom\'an and Rumbos \cite{roman-r}
tried to give a meaning to quantum operations  and Rom\'an and
Zuazua claim that Finch's condition ``is simply \it some\/ \rm
kind of \it deduction theorem\/\rm.''~\cite{roman-z} The condition really
shows a striking similarity with what one could call a \it deduction
theorem in a Boolean algebra\/\rm: $b\cap a\le c\ \Leftrightarrow\ a\le
b\to c$ [where $a\to b$, called \it standard implication\/\rm, denotes
$(a\cap b')'$] and at the same time is in an apparent striking
contradiction with a previous result by Malinowski according to which
``no consequence operation determined by any class of orthomodular
lattices admits the deduction theorem.''~\cite{malinowski}
On the other hand, there are authors who propose that we use \it quantum
operations\/ \rm as ``especially interesting from a physical
point of view.''~\cite{hoog-pyk} Many others considered particular
quantum operations according to various properties they judged
such an operation should have.$^{(8,10,15,17,25)}$
Because the whole issue is of
a general interest for characterizations of algebras underlying a
Hilbert space formulation of quantum mechanical systems, in this paper
we investigate whether it is possible to formulate orthomodular lattice
by means of non-standard ``quantum'' operations.

As we show in the next sections, it turns out that all previous authors
make use of the ordering relation $\le$
from the standardly defined ortholattice. This ordering relation
is defined by means of the standard conjunction operation:
$a\le b\>\Leftrightarrow^{\rm def}\>a\cap b=a$ and not by means of quantum
operations. However, if we wanted to switch from the
standard operations to quantum ones, we should redefine $\le$ as well.
\cite{mpqo01} In Sec.~\ref{sec:deduct} we do so and show that \it
quantum operations\/ \rm alone cannot serve us for defining quantum
logic. In Sec.~\ref{sec:oper} we construct an algebra based on the
aforementioned quantum conjunction $\cap_1$ so as to contain both a
quantum ordering and a deduction theorem, but which is not a
lattice---although one can embed an orthomodular lattice in it.
In Sections \ref{sec:oper} and \ref{sec:finite} we generalize 
the result to algebras based on quantum disjunctions $\cup_i$, 
$i=2,\dots,5$. In Sections \ref{sec:oper}, \ref{sec:qa0-5}, and 
\ref{sec:constr} we give an algebra which for both classical and 
quantum operations of quantum conjunction and disjunction have
identical structural forms for all its equations. Again, the
resulting algebra is not a lattice although one can embed an
orthomodular lattice in it.

\section{Deduction and Ordering}
\label{sec:deduct}

Let us first define an ortholattice with the help of the
lattice ordering relation as follows.~\cite{mpijtp98}

\begin{definition}\label{def:ol-standard}
An {\em ortholattice}, {\rm OL\/} is an algebra
$\langle{\cal OL}_0,',\cup,\cap\rangle$
such that the following conditions are satisfied for any
$a,b,c\in \,{\cal OL}_0$:
\begin{eqnarray}
{\rm OL1}&& a\le a\nonumber\\
{\rm OL2}&& a\le b\quad \&\quad b\le a\quad\Rightarrow\quad a=b\nonumber\\
{\rm OL3}&& a\le b\quad \&\quad b\le c\quad\Rightarrow\quad a\le c\nonumber\\
{\rm OL4}&& a\le a\cup b \quad \&\quad b\le a\cup b\nonumber\\
{\rm OL5}&& a\le c\quad \&\quad b\le c\quad\Rightarrow\quad a\cup b\le
c\nonumber\\
{\rm OL6}&& a\le b\cup b'\nonumber\\
{\rm OL7}&& a=a''\nonumber\\
{\rm OL8}&& a\le b\quad\Rightarrow\quad b'\le a'\nonumber\\
{\rm OL9}&& a\cap b=(a'\cup b')'\nonumber
\end{eqnarray}
where
\begin{eqnarray}
a\le b\ {\buildrel\rm def\over\Leftrightarrow}\ a\cup b=b,\label{eq:le}
\end{eqnarray}

In addition, since $a\cup a'=b\cup b'$ for any $a,b\in
\,{\cal OL}_0$, we define:
\begin{eqnarray}
1&{\buildrel\rm def\over=}&a\cup a'\\
0&{\buildrel\rm def\over =}&a\cap a'
\end{eqnarray}
\end{definition}

This definition will prove itself convenient and economic for
yielding our results, but since it is not widely known we
shall first show its equivalence to the following standard
definition.~\cite{beran}

\begin{lemma}The above definition is equivalent to the following
standard one.

A {\em lattice}, {\rm L\/} is an algebra
$\langle{\cal L}_0,\cup,\cap\rangle$
such that the following conditions are satisfied for any
$a,b,c\in \,{\cal L}_0$:
\begin{eqnarray}
{\rm L1a}&&a\cup b\>=\>b\cup a\nonumber\\
{\rm L1b}&&a\cap b\>=\>b\cap a\nonumber\\
{\rm L2a}&&(a\cup b)\cup c\>=\>a\cup (b\cup c)\nonumber\\
{\rm L2b}&&(a\cap b)\cap c\>=\>a\cap (b\cap c)\nonumber\\
{\rm L3a}&&a\cup (a\cap b)\>=\>a\nonumber\\
{\rm L3b}&&a\cap (a\cup b)\>=\>a\nonumber
\end{eqnarray}
\parindent=0pt
where $a\le b\ {\buildrel\rm def\over\Leftrightarrow}\ a\cup b=b$.

\parindent=20pt
An {\em ortholattice}, {\rm OL\/} is an algebra
$\langle{\cal OL}_0,\cup,\cap,',0,1\rangle$
such that $\langle{\cal OL}_0,\cup,\cap\rangle$
is a lattice and in which the unary operation $'$ and
the constants $0,1$
satisfy the following conditions for any
$a,b\in \,{\cal OL}_0$:
\begin{eqnarray}
{\rm OL1a'}&&a\cup a'\>=\>1\nonumber\\
{\rm OL1b'}&&a\cap a'\>=\>0\nonumber\\
{\rm OL2'}&&a=a''\nonumber\\
{\rm OL3'}&&a\le b\qquad\Rightarrow\qquad b'\le a'\nonumber
\end{eqnarray}
\end{lemma}

\begin{proof}That Def.~\ref{def:ol-standard} follows from
the standard definition is well-known.  In particular,
OL1 through OL5 follow from the conditions for L only.

The proof of the other direction is slightly less obvious, so,
we show how to prove L1a, L2a, and L3a.
L1b, L2b, and L3b are duals that follow using OL9 and OL7.

L1a follows from OL4, OL5, and OL2.

To prove L2a we first apply OL4 twice and from OL3 we get
$a\le(a\cup b)\cup c$, $b\le(a\cup b)\cup c$, and
$c\le(a\cup b)\cup c$.
\noindent Applying OL5 first to the last two equations
and then reapplying OL5 yields
$a\cup (b\cup c)\le(a\cup b)\cup c$\,.
Analogously, we get
$(a\cup b)\cup c\le a\cup (b\cup c)$\,.
OL2 then yields L2a.

To prove L3a we first obtain $a\cap b\le a$
from OL4, OL8, OL9, and OL7. From this, OL1 and OL5
yield $a\cup(a\cap b)\le a$.
The other direction follows from OL4, and OL2 finally
yields {\rm L3a}.
\end{proof}

Next, we introduce the following additional operations:
\begin{definition}\label{def:implications}
We define {\em quantum implications\/} as
\begin{eqnarray}
a\to_1 b&{\buildrel\rm def\over =}&
a'\cup(a\cap b)\label{eq:q-impl-1}\\
a\to_2 b&{\buildrel\rm def\over =}&
b'\to_1a'\label{eq:q-impl-2}\\
a\to_3 b&{\buildrel\rm def\over =}&
((a' \cap b)\cup(a'
\cap b'))\cup\bigl(a\cap(a' \cup b)\bigr)
\label{eq:q-impl-3}\\
a\to_4 b&{\buildrel\rm def\over =}&
b'\to_3a'\label{eq:q-impl-4}\\
a\to_5 b&{\buildrel\rm def\over =}&
((a\cap b)\cup(a' \cap b))\cup(a'
\cap b' )\label{eq:q-impl-5}
\end{eqnarray}
{\em quantum conjunctions\/} as
\begin{eqnarray}
a\cap_i b\ {\buildrel\rm def\over =}\
(a\to_i b')', \qquad i=1,\dots,5\label{eq:q-conj}
\end{eqnarray}
and {\em quantum disjunctions\/} as
\begin{eqnarray}
a\cup_i b\ {\buildrel\rm def\over =}\
a'\to_i b, \qquad i=1,\dots,5\label{eq:q-disj}
\end{eqnarray}

{\em Classical implication, disjunction}, and {\em implication}
are denoted as $a\to_0 b\>=^{\rm def}\>a'\cup b$,
$a\cup_0 b\>=^{\rm def}\>a\cup b$, and
$a\cap_0 b\>=^{\rm def}\>a\cap b$, respectively.
\end{definition}

For a subsequent use we introduce the following definition
of an orthomodular lattice.

\begin{definition}\label{def:om}An ortholattice to which any of the
following equivalent orthomodularity conditions are added:
\begin{eqnarray}
a\to_ib=1\qquad\Leftrightarrow\qquad a\le b,\qquad
i=1,\dots,5\label{eq:oml}
\end{eqnarray}
is called an {\em orthomodular lattice}, {\rm OML}.
{\rm \cite{pav87}}
\end{definition}

For the quantum operations defined above we can prove the following lemmas:

\begin{lemma}\label{lemma:q-de-m} {\em Quantum
De Morgan law} holds in any ortholattice for
quantum conjunction and disjunction:
\begin{eqnarray}
a\cap_i b=(a'\cup_i b')',\qquad i=1,\dots,5\label{eq:q-de-m}
\end{eqnarray}
\end{lemma}

\begin{lemma}\label{lemma:q-ord-r}In any {\rm OML} the
following {\em quantum
ordering relations} hold for quantum conjunctions
and disjunctions:
\begin{eqnarray}
a\le_{\cup i+}b&{\buildrel\rm def\over\Leftrightarrow}&a\cup_i
b=b,\qquad i=1,3,4,5\label{eq:q-ord-r-1+v}\\
a\le_{\cup i-} b&{\buildrel\rm def\over\Leftrightarrow}& b\cup_i
a=b,\qquad i=2,3,4,5\label{eq:q-ord-r-1-v}\\
a\le_{\cap i+} b&{\buildrel\rm def\over\Leftrightarrow}& a\cap_i
b=a,\qquad i=1,3,4,5\label{eq:q-ord-r-1-A}\\
a\le_{\cap i-} b&{\buildrel\rm def\over\Leftrightarrow}& b\cap_i
a=a,\qquad i=2,3,4,5\label{eq:q-ord-r-1+A}
\end{eqnarray}
\end{lemma}
\begin{proof}
One straightforwardly checks reflexivity, antisymmetry, and
transitivity.  In particular, if we use Lemma~\ref{lemma:le} below
(whose proof does not depend on this lemma), the proof becomes trivial.
Negative results (for $i=1,2$) follow from failures in OML MO2
(Fig.~\ref{fig:mo2o6}a).
\end{proof}

\begin{figure}[htbp]\centering
  \setlength{\unitlength}{1pt}
  \begin{picture}(240,90)(0,0)

    \put(0,0){
      \begin{picture}(90,80)(-20,-10)
        \put(40,0){\line(-2,3){20}}
        \put(40,0){\line(2,3){20}}
        \put(40,0){\line(-4,3){40}}
        \put(40,0){\line(4,3){40}}
        \put(40,60){\line(-2,-3){20}}
        \put(40,60){\line(2,-3){20}}
        \put(40,60){\line(-4,-3){40}}
        \put(40,60){\line(4,-3){40}}

        \put(40,-5){\makebox(0,0)[t]{$0$}}
        \put(25,30){\makebox(0,0)[l]{$x$}}
        \put(85,30){\makebox(0,0)[l]{$y'$}}
        \put(55,30){\makebox(0,0)[r]{$y$}}
        \put(-5,30){\makebox(0,0)[r]{$x'$}}
        \put(40,65){\makebox(0,0)[b]{$1$}}

        \put(40,0){\circle*{3}}
        \put(0,30){\circle*{3}}
        \put(20,30){\circle*{3}}
        \put(60,30){\circle*{3}}
        \put(80,30){\circle*{3}}
        \put(40,60){\circle*{3}}
      \end{picture}
    } 

    \put(200,0){
      \begin{picture}(60,80)(-10,-10)
        \put(20,0){\line(-1,1){20}}
        \put(20,0){\line(1,1){20}}
        \put(0,20){\line(0,1){20}}
        \put(40,20){\line(0,1){20}}
        \put(0,40){\line(1,1){20}}
        \put(40,40){\line(-1,1){20}}

        \put(20,-5){\makebox(0,0)[t]{$0$}}
        \put(-5,20){\makebox(0,0)[r]{$x$}}
        \put(45,20){\makebox(0,0)[l]{$y'$}}
        \put(-5,40){\makebox(0,0)[r]{$y$}}
        \put(45,40){\makebox(0,0)[l]{$x'$}}
        \put(20,65){\makebox(0,0)[b]{$1$}}

        \put(20,0){\circle*{3}}
        \put(0,20){\circle*{3}}
        \put(40,20){\circle*{3}}
        \put(0,40){\circle*{3}}
        \put(40,40){\circle*{3}}
        \put(20,60){\circle*{3}}

      \end{picture}
    } 

  \end{picture}
  \caption{\hbox to3mm{\hfill}(a)
    Lattice MO2;
  \hbox to20mm{\hfill} (b) Lattice O6.
\label{fig:mo2o6}}
\end{figure}

In what follows we sometimes use $\le_i$ to denote any of
$\le_{\cup i+}$, $\le_{\cup i-}$, $\le_{\cap i+}$, $\le_{\cap i+}$
from Lemma~\ref{lemma:q-ord-r} (excluding the exceptions for
$i=1,2$).
One can easily prove the following two lemmas, using the fact that
each condition fails in the non-orthomodular lattice O6
(Fig.~\ref{fig:mo2o6}b): \cite[p.~22]{kalmb83}

\begin{lemma}\label{lemma:le}
\begin{eqnarray}
a\le_i b\quad &\Leftrightarrow&\quad a\le b,\qquad\mbox{\rm for
all $i$ from Lemma \ref{lemma:q-ord-r}}\nonumber
\end{eqnarray}
make {\rm OL} orthomodular
and hold in any {\rm OML}.
\end{lemma}

\begin{lemma}\label{lemma:oml-vv}An ortholattice to which any of
the following conditions is added:
\begin{eqnarray}
a'\cup_ib=1&\Leftrightarrow&a\le_{\cup i+}b,
\qquad i=1,3,4,5\label{eq:oml+v}\\
a'\cup_ib=1&\Leftrightarrow&a\le_{\cup i-}b,
\qquad i=2,3,4,5\label{eq:oml-v}\\
a'\cup_ib=1&\Leftrightarrow&a\le_{\cap i+}b,
\qquad i=1,3,4,5\label{eq:oml+A}\\
a'\cup_ib=1&\Leftrightarrow&a\le_{\cap i-}b,
\qquad i=2,3,4,5\label{eq:oml-A}
\end{eqnarray}
is an {\rm OML} and vice versa.
\end{lemma}

The main theorem of this section is:

\begin{theorem}\label{th:ol-quantum}
Conditions {\rm OL1-OL3} and {\rm OL5-OL9} as well the orthomodularity
(Def.~\ref{def:om}) hold with $\le_i$ from Lemma \ref{lemma:q-ord-r}
substituted for $\le$ and with $\cup_i$ and $\cap_i$, $i=1,\dots,5$
substituted for $\cup$ and $\cap$, respectively. As for {\rm OL4} we
have:
\begin{eqnarray}
{\rm (OL4)}&& a\le_{\cap 1+}a\cup_1 b\quad\&\quad
b\nleq_{\cap 1+}a\cup_1 b\quad
\&\quad a\nleq_{\cap 2-} a\cup_2 b\quad
\&\quad b\le_{\cap 2-} a\cup_2 b\quad\&\ \nonumber\\
&&a\nleq_i a\cup_i b\quad\&\quad b\nleq_ia\cup_i b\qquad i=3,4,5.
\nonumber
\end{eqnarray}
In addition, $a\cup_i a'\>=a'\cup_i a\>=\>1$ and
$a\cap_i a'\>=a'\cap_i a\>=\>0$
hold for any $i$.
\end{theorem}
\begin{proof}
The positive results are easy to verify if we convert
$\le_i$ to $\le$ using Lemma~\ref{lemma:le}.
For OL5, we illustrate the proof for $i=3$ and $a\cup_3 b$:  from
$a\le c$ and $b\le c$, we have $a\cap b\le c$, $a\cap b'\le c$,
and $a'\cap(a\cup b)\le c$.  Hence
$(a\cap b)\cup(a\cap b')\cup(a'\cap(a\cup b))\le c$.
The negative results for OL4
are shown by their failures in OML MO2.
\end{proof}

\begin{lemma}\label{lemma:ded}The following forms of a deduction theorem
make any {\rm OL} orthomodular. (We say that $b$ is deducible from
$\Gamma$ if there exist $a_1,a_2,\dots,a_n\in\Gamma$ such that
$a_n\cap_1(a_{n-1}\cap_1(\dots a_1)\dots)\le_{1+}b$, where
$\le_{1+}$ denotes any one of $\le$,
$\le_{\cup 1+}$, $\le_{\cap 1+}$.~\cite{gold74})
\begin{eqnarray}
(b\cap_1 a)\le c\quad &\Leftrightarrow& \quad a\le b\to_1 c,
\qquad\qquad ^{(9,32)}\\
(b\cap_1 a)\le_{\cup 1+} c\quad &\Leftrightarrow& \quad a\le_{\cup 1+}
b\to_1 c,\label{eq:1-i-1}\\
(b\cap_1 a)\le_{\cap 1+} c\quad &\Leftrightarrow& \quad a\le_{\cap 1+}
b\to_1 c,\label{eq:1-i-2}
\end{eqnarray}
and these conditions hold in any orthomodular lattice, {\rm OML}.
Eqs.~(\ref{eq:1-i-1}) and (\ref{eq:1-i-2}) do not hold for any
other $i$ from Lemma \ref{lemma:q-ord-r} substituted together with the
corresponding $b\cap_ia$ for $\le_{1+}$ and $b\cap_1a$, respectively.
\end{lemma}

\smallskip\noindent
\bf Consequences of Theorem \ref{th:ol-quantum}. \rm
If we keep only to \it quantum operations\/ \rm and \it quantum
ordering\/ \rm we will arrive at an algebra which is not a lattice
(in the sense that $\cup_1$ and $\cap_1$ operations do not
coincide with supremum and infimum) because the lattice axiom OL4
is not satisfied. How this algebra can be axiomatized (starting
with those above conditions that hold in a lattice) is an open
problem. Specifically, such an axiomatization would be able to prove
those conditions that hold in any OML when $\cup$ and $\cup_1$ are
simultaneously interchanged throughout, i.e., conditions OL1-3,
valid parts of OL4, OL5-9, and the orthomodularity.
In any case the algebra cannot be made orthoisomorphic to the lattice
of closed subspaces of Hilbert space. If we still wanted
to consider an algebra based only on \it quantum operations\/ \rm
and \it quantum ordering\/ \rm we would have the following alternatives:
To start with valid conditions from Theorem
\ref{th:ol-quantum}, take them as axioms of an algebra, and add
additional axioms so as to possibly eventually arrive at a finite
axiomatization of the algebra. Or, to express
standard operations from OML by means of quantum ones and arrive
at algebras we will consider in the next Section.
Finite axiomatizability of such algebras is also an open question
(except for those studied in Section~\ref{sec:finite}).

\section{Operation Algebras QA$_{[i]}$}
\label{sec:oper}

As follows from Theorem \ref{th:ol-quantum} we cannot have a proper
``quantum'' lattice based on \it quantum operations\/\rm,
\it quantum ordering\/\rm, and equipped with
a deduction theorem since we cannot have any lattice based on
them. However, following the approach of Ref.~\citen{mpqo01}
we can arrive at the following algebra.

\begin{definition}\label{def:oml-nstandard}
Operation algebra {\rm QA$_{[i]}$} is an algebra
$\langle{\cal A}_0,',\Cup\rangle$
such that the following rule is satisfied

{\em Substitution Rule.} Any valid condition or equation
one can obtain in the standard formulation of {\rm OML}
containing only variables, $\cup_i$ (where $i$ is one, or
some, or all of $i=0,\dots,5$), and negation written
in {\rm QA$_{[i]}$} with $\Cup$ substituted for $\cup_i$ is a
valid condition or equation in {\rm QA}. Subscript $[i]$
denotes the range of $i$.  (For brevity, we omit braces
when $i$ is an $n$-tuple, so {\rm QA$_{[0,1]}$} denotes
{\rm QA$_{[\{0,1\}]}$}.)
\end{definition}

Below, by giving giving explicit lattice expressions for
$i=0$, $i=1$, and $i=0$--$5$, we implicitly define algebras
QA$_{[0]}$ (standard OML), QA$_{[1]}$ (with
$a\cup b=a\cup_1(a'\cap_1 b)$), and QA$_{[0\mbox{--}5]}$ (with
$a\cup b=a\cup_i(b\cup_i(b'\cap_i(a\cup_i(a\cap_i b')))),\
i=0,\dots,5$).

Algebra QA$_{[1]}$ does contain a deduction theorem and one
can embed orthomodular lattice in it, but it is not
a lattice
(in the sense of $\cup_1$ and $\cap_1$ coinciding with
supremum and infimum)
although it shows a striking similarity to a
classical distributive lattice. The similarity is
contained in the form of orthomodularity (Lemma
\ref{lemma:oml-vv}) and in the deduction theorem
(Lemma \ref{lemma:ded}). However, if we agree that a
deduction theorem is not a very useful theorem in a
lattice theory then Theorem \ref{th:ol-quantum} shows that
there is no particular reason to limit ourselves to $\cup_1$ for
$\Cup$. For example, in algebra QA$_{[1\mbox{--}5]}$ one can express
all relevant operations from an orthomodular lattice by means of
5 quantum disjunctions and conjunctions at once
(e.g., $a\cup b=((a\cup_ib')\cup_i(b'\cup_ia))'\cup_ia,\
i=1,\dots,5$).~\cite{mpqo01}
This might seem to take us away from any ``classical'' feature of
the algebra, but below we show that in quantum-classical algebra
QA$_{[0\mbox{--}5]}$ one can express all expressions in an
orthomodular lattice by means of either quantum or classical
conjunctions and disjunctions in structurally identical ways.
More explicitly, we express 96 possible two-variable expressions in
an orthomodular lattice (so-called Beran expressions \cite{beran})
by means of both quantum and classical disjunctions and conjunctions,
and negation.

\section{Classical-Quantum Algebra QA$_{[0\mbox{--}5]}$}\label{sec:qa0-5}

Let $n_{a,b},\ n=1,\dots,96,\ a,b\in {\rm OML}$ be a Beran expression.
\cite{beran} As we stressed in Ref.~\citen{mpqo01} there are 16 classical
and 80 quantum Beran expressions. Classical expressions can be expressed
by classical-quantum disjunctions and conjunctions (where
$\Cup=\cup_i,\ i=0,\dots,5$; DeMorgan's
law holds $a\Cap b=(a'\Cup b')'$, but we use both operations for the
compactness of expressions) and negation. Shortest such expressions
are given below. We also give shortest expressions by means of
$\cup_1$ and $\cap_1$.
\begin{eqnarray}
1&=&0'\ =\ 96_{a,b}=1_{a,b}'\ = \ 1\ =\ 0' \nonumber\\
a&=&22_{a,b}=75_{a,b}'=39_{b,a}=58_{b,a}'\ =\ a\nonumber\\
a\cup b&=&(a'\cap b')'\ =\ 92_{a,b}=93_{a,b'}=94_{a',b}=95_{a',b'}
=2_{a,b}'=3_{a,b'}'=4_{a',b}'=5_{a',b'}'\nonumber\\
&=& a\Cup(b\Cup(b'\Cap(a\Cup(a\Cap b'))))\ = \
b\cup_1(b'\cap_1a)\nonumber\\
a\equiv_0 b&=& (b\cup a')\cap (b'\cup a)\ =\ 88_{a,b}=9_{a,b}'
\nonumber\\ &=&
(a\Cup b)\Cap((a\Cap b)'\Cap((a\Cap b')\Cup(a'\Cap b)))\ =\
(b'\cap_1a')\cup_1(b\cap_1a)\nonumber
\end{eqnarray}

Quantum Beran expressions are: \it quantum unities\/\rm, $1_{abi}$,
\it quantum zeros\/\rm, $0_{abi}$, $i=1,2$, \it quantum variables\/
\rm $a_{abi}$,  $i=1,2,3$, and the above defined
\it quantum identities\/\rm, \it disjunctions\/ \rm  and \it
conjunctions\/\rm, that all reduce to
their classical counterparts for compatible variables. They can be
expressed by classical-quantum disjunctions and conjunctions
[DeMorgan's law holds $a\Cap b=(a'\Cup b')'$], and negation as
given below (shortest possible expressions). We also give their
shortest expressions by means of $\cup_1$ and
$\cap_1$.
\begin{eqnarray}
1_{ab1}&=&0_{a'b'1}'\ =\ ((a\cap b)\cup (a\cap b'))\cup
((a'\cap b)\cup (a'\cap b'))\ =\ 16_{a,b}=81_{a,b}'\nonumber\\
&=& ((a\Cup b)\Cup (b\Cup a'))\Cap(((a'\Cap b)\Cup (b\Cap a))\Cup
((a\Cup b)'\Cup (b'\Cap a)))\nonumber\\ &=&
(a\cup_1(b\cup_1a'))\cap_1(a'\cup_1(b\cup_1a))
 \nonumber\\
1_{ab2}&=&0_{a'b'2}'\ =\ (a\cup (a'\cap b))\cup (a'\cap b')
\nonumber\\ &=&
32_{a,b}=80_{a',b'}=48_{b,a}=64_{b',a'}=
17_{a',b'}'=65_{a,b}'=33_{b',a'}'=49_{b,a}'\nonumber\\ &=&
a\Cup (((b\Cup(a\Cup b)')\Cap(a\Cup (a\Cap b))')\Cup a)\ =\
a\cup_1(b\cap_1a)'\nonumber\\
a_{b1}&=&((a\cap b)\cup (a\cap b'))=6_{a,b}=7_{b,a}\!=
10_{b',a'}=11_{a',b'}=86_{a',b'}'=87_{b',a'}'=\!90_{b,a}\!'=91_{a,b}'
\nonumber\\ &=& a\Cap ((a\Cup b)\Cap (a'\Cup
((b\Cap a)\Cup (a\Cap b'))))\ =\ a\cap_1(a'\cup_1(b\cup_1a))
\nonumber\\
a_{b2}&=&(a\cup b)\cap (b'\cup (b\cap a))=54_{a,b}\!=\!\!23_{b',a}
\!=\!\!26_{b',a'}'\!=\!\!38_{a,b'}\!=\!\!43_{a',b'}'\!=\!\!59_{a,b'}'\!=\!\!
71_{b,a}\!\!=\!\!74_{b',a}'
\nonumber\\ &=& (b'\Cup (a\Cup b'))\Cap ((a\Cap b)\Cup
(((b\Cup a)\Cup a)\Cap b'))\ = \ (b'\cap_1a)\cup_1a\nonumber\\
a_{b3}&=&((a\cup b)\cap (a\cup b'))\cap ((a'\cup
(a\cap b))\cup (a\cap b'))=
70_{a,b}=27_{a',b'}=55_{b,a}=42_{b',a'}\nonumber\\ &=&
(a'\Cup (b\Cup a))\Cap (((b\Cap a)\Cup (a\Cap b'))\Cup
((a\Cup ((b\Cup a)\Cap b'))\Cap a'))\nonumber\\ &=&
(a'\cap_1(b'\cap_1a))\cup_1((b'\cap_1a)\cup_1a)
\nonumber\\
a\equiv_1 b&=&a'\equiv_3 b'\ =\ (a\cup b')\cap
(a'\cup (a\cap b))\nonumber\ =\ 72_{a,b}=
73_{a,b'}=56_{b,a}=57_{b,a'}\nonumber\\ &=& (a\Cup (b\Cup a))'\Cup
((b'\Cup a)\Cap ((b\Cap (a\Cap b))\Cup a'))\ = \
(a\cup_1b)'\cup_1(b\cap_1a) \nonumber\\
a\equiv_2 b&=&a'\equiv_4 b'\ =\ (a\cup b')\cap
(b\cup (b'\cap a'))\ = \
40_{a,b}=41_{a',b}=24_{b,a}=25_{a,b'}\nonumber\\ &=&
(b\Cap (a\Cap b))\Cup ((a\Cup b')\Cap
(b\Cup ((b\Cup a)'\Cup b)))\ = \
(b\cap_1a)\cup_1(a\cup_1(b\cup_1a))' \nonumber\\
a\equiv_5 b&=&(b'\cap a')\cup (b\cap a)\ =\
8_{a,b}=89_{a,b}'\nonumber\\ &=&
(a\Cup b')\Cap ((a\Cup (b\Cup a))'\Cup ((b\Cap a)\Cap b))
\ =\ (b\cup_1a')\cap_1(b'\cup_1a) \nonumber\\
a\cup_1b&=& b\cup_2a\ =\ (a'\cap_1b')'\ =\ (b'\cap_2a')'
\ = \ a\cup (a'\cap b)\nonumber\\
&=&28_{a,b}=29_{a,b'}=44_{b,a}=46_{b,a'}=61_{b',a}=
63_{b',a'}=78_{a',b}=79_{a',b'}\nonumber\\
&=&18_{a',b'}'=19_{a',b}'=34_{b',a'}'=36_{b',a}'=51_{b,a'}'=
53_{b,a}'=68_{a,b'}'=69_{a,b}'\nonumber\\
&=&(a\Cup (b\Cap(a\Cup (a\Cap b))'))\nonumber\\
a\cup_3b&=& b\cup_4a\ =\ (a'\cap_3b')'\ =\ (b'\cap_4a')'\ =\
(a\cup b)\cap ((a'\cup (a\cap
b'))\cup (a\cap b))\nonumber\\ &=& 76_{a,b}=30_{a',b}=
31_{a',b'}=45_{b',a}=47_{b',a'}=60_{b,a}=62_{b,a'}=77_{a,b'}
\nonumber\\
&=&66_{a',b'}'=20_{a,b'}'=21_{a,b}'=35_{b',a}'=37_{b,a}'=
50_{b',a'}'=52_{b',a}'=67_{a',b}'\nonumber\\
&=&(a'\Cup (b\Cup a))\Cap (((b\Cap a)\Cup (a\Cap b'))\Cup
((a\Cup b)\Cap a'))\ = \ (a'\cap_1b)\cup_1(b\cup_1a)\nonumber\\
a\cup_5b&=& (a'\cap_5b')'\ = \
((a\cap b)\cup (a\cap b'))\cup (a'\cap b)
\nonumber\\ &=& 12_{a,b}=13_{a',b}=14_{a',b'}=15_{a,b'}=
82_{a',b'}'=83_{a',b}'=84_{a,b}'=85_{a,b'}'\nonumber\\ &=&
(a\Cup b)\Cap (((a'\Cap b)\Cap (b\Cup a))\Cup
((b\Cap a)\Cup (a\Cap b')))\nonumber\\ &=&
(a\cup_1b)\cap_1(a'\cup_1(b'\cup_1a))\nonumber
\end{eqnarray}

Each of the 96 $\Cup,\Cap$ expressions above (which are polynomials in
$\Cup,\Cap,',0,1$) is a shortest representative of an (infinite)
equivalence class of polynomials of QA$_{[0\mbox{--}5]}$ that evaluate (in an
OML) to the same Beran expression for each of $\cup_i,\cap_i,
i=0,\ldots,5$ substituted for $\Cup,\Cap$.  In addition to these
polynomials, there are many other polynomials that do not evaluate to
the same Beran expression for all $i$, for example $a\Cup b$.  Every
polynomial belongs to an equivalence class of polynomials
that can be connected with the $=$ sign.  For example, $a\cup_i b=
a\cup_i(b\cup_i(a\cap_i(a\cup_i b)))$ holds for $i=0,\ldots,5$, so
$a\Cup(b\Cup(a\Cap(a\Cup b)))$ belongs to the same equivalence class as
$a\Cup b$.  In fact there are $2^4 \times 6^6 = 746496$ such equivalence
classes of expressions with at most 2 distinct variables.

By means of an exhaustive computer search, the authors determined that a
shortest polynomial for a member of each equivalence class requires from
0 to 14 variable occurrences.  We have $0$ and $1$ with no variable
occurrences; $a, b, a', b'$ with one occurrence; $a\Cup b, b\Cup a,
a\Cap b'$, etc.\ (16 cases) with two occurrences; and so on.  An example
of a shortest representative of a class requiring at least 14 variable
occurrences is $(a\Cap (b\Cap (b\Cup a)))\Cup ((a'\Cap ((b\Cup a)\Cap
b))\Cup ((b\Cup ((b\Cup (a\Cup b))\Cap a'))\Cap b'))$.  The number of
equivalence classes vs.\ the number of variable occurrences required for
a shortest representative is shown in Table~\ref{tab:equivclasses}.  The
computer search yielded a list, available from the authors, with a
shortest representative of each of the 746496 classes.  (The list also
includes as subsets shortest representatives for all possible operation
algebras QA$_{[i]}$, $i\subseteq\{0,\ldots,5\}$. The next section shows
how these can be extracted from the list.)

\begin{table}[hbtp]
{\small\begin{center}
\begin{tabular}{|c|c|c|c|c|c|c|c|c|c|c|c|c|c|c|c|}   \hline
(a) &

$\!\!\!\!$  2 $\!\!\!\!$      &
$\!\!\!\!$   4   $\!\!\!\!$    &
$\!\!\!\!$   16  $\!\!\!\!$    &
$\!\!\!\!$   224   $\!\!\!\!$  &
$\!\!\!\!$   1926  $\!\!\!\!$  &
$\!\!\!\!$   10568 $\!\!\!\!$  &
$\!\!\!$   29444 $\!\!\!\!$  &
$\!\!\!$   101168$\!\!$  &
$\!\!\!$   195380 $\!\!\!$ &
$\!\!\!$   204296$\!\!$  &
$\!\!\!$   138584$\!\!$  &
$\!\!\!$    48852$\!\!$  &
$\!\!\!\!$    14272$\!\!$  &
$\!\!\!\!$    1684 $\!\!\!\!$  &
$\!\!\!\!$    76  $\!\!\!\!$   \\
(b)
  & 0
  & 1
  & 2
  & 3
  & 4
  & 5
  & 6
  & 7
  & 8
  & 9
  & 10
  & 11
  & 12
  & 13
  & 14\\ \hline
\end{tabular}
\end{center}}
  \caption{(a)
    Number of equivalence classes for QA$_{[0\mbox{--}5]}$ polynomials 
    with at most 2 distinct variables (total 746496);
   (b) number of variable occurrences in a shortest representative
    of those classes.
\label{tab:equivclasses}}
\end{table}

There are usually several shortest representatives of a given equivalence
class, and the relationship is not always obvious.  For example,
$a\Cap(b\Cup a)'=a\Cap(b\Cap a')$ are two shortest representatives
of one class.  Other interesting classes include those that can
permute quantum operations to others; for example
$((b'\Cap a)\Cap (a\Cup b))\Cup ((a\Cap b)\Cup (b\Cap a'))$ i.e.\
$((b'\cap_i a)\cap_i (a\cup_i b))\cup_i ((a\cap_i b)\cup_i (b\cap_i a'))$
evaluates to $a\cup_{5-i}b, i=0,\ldots,5$.

The situation for $\Cup,\Cap$ expressions with 3 or more distinct
variables appears to be much more complicated and is poorly understood.
Apparently the only known non-trivial conditions with 3 variables in
QA$_{[0\mbox{--}5]}$ are the Foulis-Holland-like associativity and
distributivity properties presented in Ref.~\citen{mpqo01}, the condition 
OL5 of Theorem~\ref{th:ol-quantum}, and the following conditions for
the Sasaki projection $\varphi_a(b)  \>=^{\rm def}\>  a\cap(a'\cup b)
= a\Cap (b\Cup (a\Cap (a\Cup b))')$.
\begin{theorem}
The following conditions hold in {\rm QA$_{[0\mbox{--}5]}$}:
\begin{eqnarray}
\varphi_a(c)=\varphi_b(c)\quad & \Rightarrow & \quad 
a\Cap c=b\Cap c\label{eq:sasa}\\
\varphi_a(c')=\varphi_b(c')\quad & \Rightarrow & \quad 
a\Cup c=b\Cup c\label{eq:sasb}
\end{eqnarray}
\end{theorem}
\begin{proof}
For (\ref{eq:sasa}), we successively
operate on both sides of the hypothesis with either
$c$ or a previous equality, using the identities
$a\cap c=c\cap \varphi_a(c)$,
$a\cap_1 c=\varphi_a(c)$,
$a\cap_2 c=(c\to_1 (\varphi_a(c))')'$,
$a\cap_3 c=(((a\cap_1 c)\to_1 c)\to_1(a\cap_2 c)')'$,
$a\cap_4 c=((c\to_1 (a\cap_1 c))\to_1 (a\cap_1 c)')'$,
and $a\cap_5 c=(a\cap_1 c)\cup (a\cap_2 c)$ on
the left-hand side (and the analogous ones for $b$
on the right-hand side).
For (\ref{eq:sasb}),
we use the identity $\varphi_{a'}(c')=(c'\to_2 \varphi_a(c'))'$ after
operating
on both sides with $c$, then use (\ref{eq:sasa}) to obtain
$a'\Cap c'=b'\Cap c'$, from which the result follows by complementing
both sides and applying DeMorgan's law.
\end{proof}
\noindent {\em Comment.}
{}From the identity $\varphi_{a'}(c)=(c\to_2 \varphi_a(c))'$, we also note
the interesting consequence $\varphi_{a}(c)=\varphi_{b}(c)
\Leftrightarrow \varphi_{a'}(c)=\varphi_{b'}(c)$, which holds in all
OMLs and is equivalent to the orthomodular law.
This is also equivalent to  
$a\to_1c=b\to_1c\ \Leftrightarrow\ a'\to_1c=b'\to_1c$.     

An open problem is whether QA$_{[0\mbox{--}5]}$ can be represented with
a finite set of equations.

\section{Construction of QA$_{[0\mbox{--}5]}$ Expressions}
\label{sec:constr}

In the previous section we showed several examples of shortest
representatives of the the 746496 equivalence classes for 2-variable
QA$_{[0\mbox{--}5]}$ expressions, which were found by exhaustive search.  
In this section we describe an algorithm that will produce a
representative of any desired equivalence class.  Although the
expressions produced with this algorithm will usually be extremely long
and not practical to work with, the algorithm is nonetheless instructive
as it illustrates another way to describe and classify the equivalence
classes.  It also provides a proof that all possible 746496
equivalence classes can be represented in QA$_{[0\mbox{--}5]}$.

The 96 Beran expressions correspond to the 96 elements of the free OML
$F(a,b)$ with 2 free generators $a,b$.  Each element can be separated
into a ``Boolean part'' and an ``MO2 part.''  \cite{navara} A Beran
expression can be identified with an ordered pair of numbers, the first
of which (0 through 15) identifies the Boolean part
$n_{\mbox{\scriptsize{}B}}$ and the second (0 through 5) the MO2 part
$n_{\mbox{\scriptsize{}M}}$.  We choose these numbers in such a way that
Beran's numbering $n_{\mbox{\scriptsize{}Beran}}$ (1 through 96) is
recovered by $n_{\mbox{\scriptsize{}Beran}}=16 n_{\mbox{\scriptsize{}B}}
+ n_{\mbox{\scriptsize{}M}} + 1$.  Conversely, given the Beran number
the Boolean part is
$n_{\mbox{\scriptsize{}B}}=(n_{\mbox{\scriptsize{}Beran}}-1)\bmod 16$
and the MO2 part is $n_{\mbox{\scriptsize{}M}}=\lfloor
(n_{\mbox{\scriptsize{}Beran}}-1)/16 \rfloor$.

\begin{table}[hbtp]
{\small\begin{center}
\begin{tabular}{|c|c|}   \hline
$\langle 0,5,5,5,5,5,5\rangle$ & $((a\Cap b)\Cap (b\Cap a'))\Cup (((a'\Cup
  b)\Cap (b\Cup a))\Cap ((a\Cap b)'\Cap (b'\Cup a)))$\\
$\langle 1,5,5,5,5,5,5\rangle$ & $(a\Cap b)\Cup (((a'\Cup b)\Cap (b\Cup
  a))\Cap ((b\Cup a)\Cap (a\Cup b')))$\\
$\langle 2,5,5,5,5,5,5\rangle$ & $(a\Cap b')\Cup (((a\Cap b)'\Cap (b\Cup
  a))\Cap ((b'\Cup a)\Cap (a\Cup b)))$\\
$\langle 3,5,5,5,5,5,5\rangle$ & $(b\Cap a')\Cup (((b\Cap a)'\Cap (a\Cup
  b))\Cap ((a'\Cup b)\Cap (b\Cup a)))$\\
$\langle 4,5,5,5,5,5,5\rangle$ & $((a\Cup b)\Cap (((a'\Cap b)\Cap (b\Cup
  a))\Cup ((b\Cap a)\Cup (a\Cap b'))))'$\\
$\langle 5,5,5,5,5,5,5\rangle$ & $a\Cup ((a\Cap b)\Cup (a'\Cap ((b\Cup
  a)\Cap (a\Cup b'))))$\\
$\langle 6,5,5,5,5,5,5\rangle$ & $b\Cup ((a\Cap b)\Cup (a'\Cap ((b\Cup
  a)\Cap (a\Cup b'))))$\\
$\langle 7,5,5,5,5,5,5\rangle$ & $(a\Cap b)\Cup ((a\Cup b)'\Cup ((a\Cup
  b')\Cap (a'\Cup b)))$\\
$\langle 8,5,5,5,5,5,5\rangle$ & $(a\Cap b')\Cup ((a\Cap (b\Cap a))'\Cap
  ((b\Cup a)\Cup b))$\\
$\langle 9,5,5,5,5,5,5\rangle$ & $b'\Cup ((a\Cup b)'\Cup (a\Cap ((b\Cap
  a)'\Cap (a\Cup b))))$\\
$\langle 10,5,5,5,5,5,5\rangle$ & $a'\Cup ((a\Cup b)'\Cup (((a\Cup b)\Cap
  (b\Cap a)')\Cap b))$\\
$\langle 11,5,5,5,5,5,5\rangle$ & $a\Cup (b\Cup (b'\Cap (a\Cup (a\Cap b'))))$\\
$\langle 12,5,5,5,5,5,5\rangle$ & $a\Cup (b'\Cup (b\Cap (a\Cup (a\Cap b))))$\\
$\langle 13,5,5,5,5,5,5\rangle$ & $b\Cup (a'\Cup (a\Cap (b\Cup (b\Cap a))))$\\
$\langle 14,5,5,5,5,5,5\rangle$ & $(a\Cap (b'\Cap a))\Cup (a\Cap (b\Cap a))'$\\
$\langle 15,5,5,5,5,5,5\rangle$ & $1$\\
\hline
\end{tabular}
\end{center}}
  \caption{Equivalence classes and shortest representatives for the
  Boolean part of a QA$_{[0\mbox{--}5]}$ construction.
These correspond to Beran expressions 81 through 96, respectively.
\label{tab:equivbool}}
\end{table}

\begin{table}[hbtp]
{\small\begin{center}
\begin{tabular}{|c|c|}   \hline
$\langle 15,0,5,5,5,5,5\rangle$ & $(a\Cap b)\Cup ((a\Cap b')\Cup
  (a'\Cap ((b\Cup a)'\Cup b)))$\\
$\langle 15,1,5,5,5,5,5\rangle$ & $a\Cup ((a'\Cap b)\Cup
(b\Cup a)')$\\
$\langle 15,2,5,5,5,5,5\rangle$ & $b\Cup ((b'\Cap a)\Cup
(a\Cup b)')$\\
$\langle 15,3,5,5,5,5,5\rangle$ & $b'\Cup ((b\Cap a)\Cup
(a'\Cap b))$\\
$\langle 15,4,5,5,5,5,5\rangle$ & $a'\Cup ((a\Cap b)\Cup
(b'\Cap a))$\\
\hline
$\langle 15,5,0,5,5,5,5\rangle$ & $(a\Cap (a'\Cup b))\Cup
((a\Cup (b\Cup a'))\Cap ((a\Cap b)'\Cup b))$\\
$\langle 15,5,1,5,5,5,5\rangle$ & $a\Cup (a\Cup ((b\Cup a)\Cup a'))$\\
$\langle 15,5,2,5,5,5,5\rangle$ & $b\Cup (a\Cup (b\Cap (b\Cup a))')$\\
$\langle 15,5,3,5,5,5,5\rangle$ & $b'\Cup (a\Cup ((b\Cap a)\Cup b))$\\
$\langle 15,5,4,5,5,5,5\rangle$ & $a'\Cup (b\Cup ((a\Cap b)\Cup a))$\\
\hline
$\langle 15,5,5,0,5,5,5\rangle$ & $((a\Cup b)\Cap b')\Cup ((a\Cup
(a'\Cup b))\Cap ((b\Cup a)\Cup b'))$\\
$\langle 15,5,5,1,5,5,5\rangle$ & $a\Cup (b\Cup (b'\Cup (b\Cup a)))$\\
$\langle 15,5,5,2,5,5,5\rangle$ & $a\Cup (a'\Cup ((a\Cup b)\Cap b))$\\
$\langle 15,5,5,3,5,5,5\rangle$ & $a\Cup (a\Cap ((a\Cup b)\Cap b))'$\\
$\langle 15,5,5,4,5,5,5\rangle$ & $b\Cup (b\Cap ((b\Cup a)\Cap a))'$\\
\hline
$\langle 15,5,5,5,0,5,5\rangle$ & $(a\Cap (a\Cup (b\Cup a)))\Cup
(a\Cup
  (a\Cap (b\Cup a)))'$\\
$\langle 15,5,5,5,1,5,5\rangle$ & $a\Cup (a'\Cup (b\Cup a))$\\
$\langle 15,5,5,5,2,5,5\rangle$ & $b\Cup (b'\Cup (a\Cup b))$\\
$\langle 15,5,5,5,3,5,5\rangle$ & $b\Cup (a\Cup (a\Cup b)')$\\
$\langle 15,5,5,5,4,5,5\rangle$ & $a\Cup (b\Cup (b\Cup a)')$\\
\hline
$\langle 15,5,5,5,5,0,5\rangle$ & $(a\Cap ((a\Cup b)\Cup a))\Cup
(a\Cup((a\Cup b)\Cap a))'$\\
$\langle 15,5,5,5,5,1,5\rangle$ & $a\Cup ((a\Cup b)\Cup a')$\\
$\langle 15,5,5,5,5,2,5\rangle$ & $b\Cup ((b\Cup a)\Cup b')$\\
$\langle 15,5,5,5,5,3,5\rangle$ & $a\Cup ((a\Cup b)'\Cup b)$\\
$\langle 15,5,5,5,5,4,5\rangle$ & $b\Cup ((b\Cup a)'\Cup a)$\\
\hline
$\langle 15,5,5,5,5,5,0\rangle$ & $(a\Cup b)\Cup (a\Cap b)'$\\
$\langle 15,5,5,5,5,5,1\rangle$ & $a\Cup (a\Cup (b\Cup (b\Cap a)'))$\\
$\langle 15,5,5,5,5,5,2\rangle$ & $b\Cup (a\Cup ((a\Cap b)'\Cup b))$\\
$\langle 15,5,5,5,5,5,3\rangle$ & $b'\Cup (b\Cup (a\Cup (a\Cup b)))$\\
$\langle 15,5,5,5,5,5,4\rangle$ & $a'\Cup (a\Cup (b\Cup (b\Cup a)))$\\
\hline
\end{tabular}
\end{center}}
  \caption{Equivalence classes and shortest representatives for the
  MO2 parts of a QA$_{[0\mbox{--}5]}$ construction.  
  These have no corresponding Beran expressions.
\label{tab:equivmo2}}
\end{table}

The Boolean part of any expression of QA$_{[0\mbox{--}5]}$ is the same for
$i=0,\ldots,5$ when $\cup_i,\cap_i$ are substituted for $\Cup,\Cap$.
The MO2 part can differ for each $i$.  We represent any equivalence
class by a septuple $\langle n_{\mbox{\scriptsize{}B}},
n_{{\mbox{\scriptsize{}M}}_0},\ldots, n_{{\mbox{\scriptsize{}M}}_5}
\rangle$ where $n_{\mbox{\scriptsize{}B}}$ identifies the Boolean part
and $n_{{\mbox{\scriptsize{}M}}_0},\ldots,
n_{{\mbox{\scriptsize{}M}}_5}$ the MO2 parts.

When
$n_{{\mbox{\scriptsize{}M}}_0}=\cdots=n_{{\mbox{\scriptsize{}M}}_5}$,
the expression is in the equivalence class of one of the 96 Beran
expressions.  For example, $a\cup b= a\Cup(b\Cup(b'\Cap(a\Cup(a\Cap
b'))))$ from the previous section, with Beran number 92, is in the
equivalence class $\langle 11,5,5,5,5,5,5 \rangle$.  When the
$n_{{\mbox{\scriptsize{}M}}_i}$ are not all the same, the equivalence
class does not correspond to any Beran expression.  For example,
$a\Cup b$ is in the equivalence class $\langle 11,5,1,2,4,3,0
\rangle$. The example $((b'\Cap a)\Cap (a\Cup b))\Cup ((a\Cap b)\Cup
(b\Cap a'))$ from
the previous section is in the equivalence class $\langle 11,0,3,4,2,1,5
\rangle$, allowing one to easily see that it evaluates to $a\cup_{5-i}b,
i=0,\ldots,5$ by reversing the order of the six MO2 components.
Continuing with some other examples from the previous section, we have
$\langle 0,0,0,0,0,0,0\rangle$ for $0$, $\langle 15,5,5,5,5,5,5\rangle$
for $1$, $\langle 5,1,1,1,1,1,1\rangle$ for $a$, $\langle 6,2,2,2,2,2,2
\rangle$ for $b$, $\langle 10,4,4,4,4,4,4\rangle$ for $a'$, $\langle
9,3,3,3,3,3,3\rangle$ for $b'$,$\langle 1,0,1,2,4,3,5\rangle$ for $a\Cap
b$, $\langle 1,0,2,1,3,4,5\rangle$ for $b\Cap a$,
$\langle 0,0,1,0,4,1,1
\rangle$ for $a\Cap(b\Cap a')$,
and $\langle 6,3,5,5,0,3,3 \rangle$ for
$(a\Cap (b\Cap (b\Cup a)))\Cup ((a'\Cap ((b\Cup a)\Cap b))\Cup ((b\Cup
((b\Cup (a\Cup b))\Cap a'))\Cap b'))$.

For operation algebras other than QA$_{[0\mbox{--}5]}$, a representative
expression can be obtained by simply ignoring the omitted MO2 component(s).
For example, to obtain a shortest representative for $a\cup b$ in
QA$_{[1]}$, we look at each septuple of the form $\langle
11,.,5,.,.,.,.\rangle$ where ``$.$'' means ``don't care,'' and pick a
shortest from the list of 746496 mentioned in the previous section.  In
this case a shortest is $a\cup b=a\cup_1(a'\cap_1 b)$, obtained from
$a\Cup(a'\Cap b)$ in equivalence
class $\langle 11,1,5,2,1,2,5\rangle$.  There are also 3 other
shortest ones:
$b\Cup(b'\Cap a)$ in $\langle 11,2,5,1,2,1,5\rangle$,
$(a'\Cap b)\Cup a$ in $\langle 11,1,5,1,1,4,5\rangle$, and
$(b'\Cap a)\Cup b$ in $\langle 11,2,5,2,2,3,5\rangle$.

Now we are ready to describe an algorithm for constructing a
representative expression for a given septuple $\langle
n_{\mbox{\scriptsize{}B}}, n_{{\mbox{\scriptsize{}M}}_0},\ldots,
n_{{\mbox{\scriptsize{}M}}_5} \rangle$.  In what follows, for brevity we
will interchangeably use expressions and their corresponding septuples.
Our construction starts with the Boolean part chosen from
Table~\ref{tab:equivbool}, obtaining $\langle n_{\mbox{\scriptsize{}B}},
5,5,5,5,5,5 \rangle$.  We then operate on each of the MO2 components in
succession, changing them from 5 to the desired value.

For the MO2 components that we want to keep unchanged, we make use of
the fact that $a\Cap 1=a$ for any $a$.  More specifically, the expression 
$\langle n_{\mbox{\scriptsize{}B}}, \ldots, 
n_{{\mbox{\scriptsize{}M}}_i},\ldots
\rangle \Cap \langle 15,\ldots,5,\ldots \rangle$ is equal to some
other expression $\langle n_{\mbox{\scriptsize{}B}}, \ldots,
n_{{\mbox{\scriptsize{}M}}_i},\ldots \rangle$ where MO2 component
$n_{{\mbox{\scriptsize{}M}}_i}$ is unchanged but the other MO2 components
are possibly different.

The expressions from Table~\ref{tab:equivmo2} allow us to operate on
specific MO2 components in succession while keeping all other MO2
components unchanged.  (The table omits the cases where
$n_{{\mbox{\scriptsize{}M}}_i}=5$ since in those cases there is nothing
to be done.)  To construct the MO2 part for $i=0$, we pick $\langle 15,
n_{{\mbox{\scriptsize{}M}}_0},5,5,5,5,5 \rangle$ from the first part of
Table~\ref{tab:equivmo2} and obtain \begin{eqnarray} \langle
n_{\mbox{\scriptsize{}B}}, 5,5,5,5,5,5 \rangle \Cap \langle 15,
n_{{\mbox{\scriptsize{}M}}_0},5,5,5,5,5 \rangle &=& \langle
n_{\mbox{\scriptsize{}B}}, n_{{\mbox{\scriptsize{}M}}_0},5,5,5,5,5
\rangle\,.  \end{eqnarray} To construct the MO2 part for $i=1$, we pick
$\langle 15, 5,n_{{\mbox{\scriptsize{}M}}_1},5,5,5,5 \rangle$ from the
second part of Table~\ref{tab:equivmo2} and obtain \begin{eqnarray}
\langle n_{\mbox{\scriptsize{}B}},
n_{{\mbox{\scriptsize{}M}}_0},5,5,5,5,5 \rangle \Cap \langle 15,
5,n_{{\mbox{\scriptsize{}M}}_1},5,5,5,5 \rangle &=& \langle
n_{\mbox{\scriptsize{}B}},
n_{{\mbox{\scriptsize{}M}}_0},n_{{\mbox{\scriptsize{}M}}_1},5,5,5,5
\rangle\,.  \end{eqnarray} Continuing in this fashion for $i=2,3,4,5$,
we finally arrive at an expression representing $\langle
n_{\mbox{\scriptsize{}B}}, n_{{\mbox{\scriptsize{}M}}_0},\ldots,
n_{{\mbox{\scriptsize{}M}}_5} \rangle$.

\section{Finite Axiomatizations for QA$_{[0]},\ldots,$QA$_{[5]}$}
\label{sec:finite}

As we mentioned above, it is an open problem whether in general
QA$_{[i]}$ is finitely axiomatizable in general for $i\subseteq
\{0,\ldots,5\}$.  However, when $i$ is a singleton, finite axiomatizations
exist.  QA$_{[0]}$ is of course just standard OML.  We show them
for each singleton in what follows by displaying a specific finite
axiomatization.

For the following definition we stress that $\cup$ is a {\em defined} (not
primitive) operation.

\begin{definition}\label{def:qa-prime}
An algebra {\rm QA$_{[i]}^{'}$} (where $i$ is one of $0,\ldots,5$) is a 
triple $\langle{\cal A}_i,',\Cup\rangle$ satisfying axioms
$\mbox{\rm QA}1_i^{'}$---$\mbox{\rm QA}10_i^{'}$, where
$\mbox{\rm QA}1_i^{'}$---$\mbox{\rm QA}8_i^{'}$ are displayed identically
to axioms {\rm OL1}---{\rm OL8} of Definition~\ref{def:ol-standard} and
for any
$a,b\in \,{\cal A}_i$:
\begin{eqnarray}
\mbox{\rm QA}9_i^{'}&& a\le b\quad\Rightarrow\quad a\cup(a'\cap b)=b
\nonumber\\
\mbox{\rm QA}10_i^{'}&& a\Cup b=a\cup_ib,\nonumber 
\end{eqnarray}
where
\begin{eqnarray}
a\Cap b&{\buildrel\rm def\over=}&(a'\Cup b')'\nonumber\\
a\cup b&{\buildrel\rm def\over=}&a\Cup(b\Cup(b'\Cap(a\Cup(a
\Cap b'))))\nonumber\\
a\cap b&{\buildrel\rm def\over=}&(a'\cup b')'\nonumber\\
a\le b&{\buildrel\rm def\over\Leftrightarrow}& a\cup b=b,\nonumber
\end{eqnarray}
and where $\cup_i$ is defined as in Def.~\ref{def:implications} 
only with the above defined $\cup$ and $\cap$ which are 
themselves defined by means of the primitive $\Cup$. 
\end{definition}

\begin{theorem}\label{th:finite-qa}
Algebra {\rm QA$_{[i]}^{'}$} is an axiomatization for (i.e.\ is 
equivalent to) algebra {\rm QA$_{[i]}$} from 
Definition~\ref{def:oml-nstandard} where $i$ is one of $0,\ldots,5$.
\end{theorem}
\begin{proof}
It is straightforward to show that each of 
$\mbox{\rm QA}1_i^{'}$--$\mbox{\rm
QA}10_i^{'}$ holds in an OML when $\Cup$ is replaced with
$\cup_i$ and thus is an axiom of {\rm QA$_{[i]}$}
according to the Substitution Rule.

For the converse, we note that the axioms $\mbox{\rm
QA}1_i^{'}$--$\mbox{\rm QA}9_i^{'}$ are {\em structurally} identical to
the axioms for an OML (compare Definition~\ref{def:ol-standard} to
$\mbox{\rm QA}1_i^{'}$--$\mbox{\rm QA}8_i^{'}$, using the definition of
$\cap$ for the structure of OL9, and adding $\mbox{\rm QA}9_i^{'}$ for
the structure of the OML law).  With this structure, we can mimic an OML
proof of any condition (equation or inference) involving (in addition to
$'$) only the defined symbol $\cup$.

Suppose $E$ is an axiom of {\rm QA$_{[i]}$} per the Substitution Rule.
We write down a condition $E'$ with only $\cup$ symbols, obtained from
$E$ by expanding all $\Cup$ symbols according to the right-hand-side of
$\mbox{\rm QA}10_i^{'}$.  Then $E'$ will be structurally identical to a
condition that holds in OML (because $\mbox{\rm QA}10_i^{'}$ is
structurally identical to the OML definition for $\cup_i$), so $E'$ can
be proved using $\mbox{\rm QA}1_i^{'}$--$\mbox{\rm QA}9_i^{'}$ to mimic
the OML proof.  Then from $E'$ we obtain $E$ by applying $\mbox{\rm
QA}10_i^{'}$.  \end{proof}

The axioms for QA$_{[i]}^{'}$ can be quite long when expressed in terms
of the primitive $\Cup$.  Some economy can be achieved by replacing the
definition of $\cup$ with 6 different ones, one for each $i$, by making
use of the OML identities $a\cup b= a\cup_0 b= a\cup_1(a'\cap_1 b)=
b\cup_2(b'\cap_2 a)= a\cup_3(a\cup_3 b)= a\cup_4(b\cup_4 a)
=a\cup_5(a'\cap_5 b)$.  In addition, QA$_{[0]}^{'}$ reduces to standard
OML, and for QA$_{[1]}^{'}$ an implicational algebra such as the one in
Ref.~\citen{abbott76} might provide a starting point for a more compact 
system.

For singleton $i$, it is easy to see that each {\rm QA$_{[i]}$} is
equivalent to an OML system as the next theorem shows.

\begin{theorem}
Algebras {\rm QA$_{[i]}^{'}$}, where $i$ is one of $0,\ldots,5$, are
equivalent to an {\rm OML} (and thus to each other and to
each {\rm QA$_{[i]}$} as well).
\end{theorem}
\begin{proof}
For one direction, we define $\cup$ in terms of $\Cup$ per
Defintion~\ref{def:qa-prime}.  For the other direction, we define $\Cup$
in terms of $\cup$ by treating $\mbox{\rm QA}10_i^{'}$ as a definition.
That the axioms of each system are satisfied in the other is a
straightforward verification.
\end{proof}

For algebras such as QA$_{[0\mbox{--}5]}$ with non-singleton $i$, 
the equivalence in the above sense does not hold.  For, although we can 
define $a\cup b$ in terms of $a\Cup b$, we cannot define $a\Cup b$ in 
terms of $a\cup b$. With the first definition, we can write down axioms 
from the Substitution Rule that correspond to the axioms for an OML, 
but there are additional axioms such as 
$a\Cap(b\Cup a)'=a\Cap(b\Cap a')$ that have
no corresponding $\cup$ version under this embedding.

\section{Conclusion}
\label{sec:concl}

Our results have shown that the so-called \it quantum
operations\/ \rm cannot be used for building an operation algebra
underlying Hilbert space. In particular, Theorem \ref{th:ol-quantum}
proves that no quantum operation can satisfy all lattice
conditions. Hence, the aforementioned operation algebra cannot have
the lattice of closed subspaces of Hilbert space as its model.
This is not a problem though, because in Hilbert  lattice
the standard conjunction $a\cap b$, corresponds to
set intersection, ${\cal H}_a\bigcap{\cal H}_b$, of subspaces ${\cal
H}_a,{\cal H}_b$ of Hilbert space ${\cal H}$, the ordering relation
$a\le b$ corresponds to ${\cal H}_a\subseteq{\cal H}_b$, the
standard disjunction  $a\cup b$, corresponds to the smallest closed
subspace of $\cal H$ containing ${\cal H}_a\bigcup{\cal H}_b$, and
$a'$ corresponds to ${\cal H}_a^\perp$, the set of vectors orthogonal
to all vectors in ${\cal H}_a$. We then prove that Hilbert lattice
is orthoisomorphic to the lattice of closed subspaces of any Hilbert
space and this is what we need Hilbert lattice for. From the framework
of quantum theory and/or Hilbert space theory no need for new
algebras based on  \it quantum operations\/ \rm emerges.

The whole issue of quantum operations have recently been put forward
by a reconsideration of the so-called \it deduction theorem\/
\rm within a lattice theory originally formulated by Finch
30 years ago.\cite{finch70} Rom\'an and Zuazua \cite{roman-z}
attempted to give a new formal and Hilbert space interpretation of
Finch's formulation. Essentially they advocated the
usage of the quantum operations the theorem is based on, instead of
the standard lattice operations. Other authors also recently
advocated their usage  as ``especially interesting from a physical
point of view.''~\cite{hoog-pyk} However, they all failed to substitute
quantum for standard operations in the definition of the ordering
relation too, and this is what we did in Sec.~\ref{sec:deduct}.
Our Theorem \ref{th:ol-quantum} shows that a consistent substitution
of quantum for standard operations take us to algebras none of which
is a lattice and for which we do not know whether they are all
finitely axiomatizable. To be more specific, when we use only
quantum operations---even to define the ordering relation---then
it turns out that the set of equations we get, satisfies all but one
conditions of an orthomodular lattice. Therefore, we can try to
use the conditions which are satisfied by quantum operations
according to Theorem \ref{th:ol-quantum} to eventually arrive
at a hypothetical well defined algebraic system which might even be
finitely axiomatizable. Or, we can disregard these conditions altogether
and simply express all standard operations by means of quantum
ones. This is what we did in Sec.~\ref{sec:oper}. We arrive at
well defined algebras which we call \it operation algebras\/ \rm
in analogy to implication algebras.$^{(1,28)}$
Some of them are finitely axiomatizable (``singleton'' operation
algebras from Sec.~\ref{sec:finite}) and for the others this is
still an open question (see Sec.~\ref{sec:qa0-5}). All operation
algebras properly contain any orthomodular lattice and 
``singleton'' operation algebras are moreover completely 
equivalent to OML and to each other as we show in 
Sec.~\ref{sec:finite}. Still, none of these algebras \it is\/ 
\rm a lattice (in the sense of satisfying condition OL4 of 
Theorem \ref{th:ol-quantum}) and therefore they can hardly play 
any role in the quantum theory from a foundational point of view.

However, it is not a problem but a virtue of quantum logic
that the only way to formulate it as an orthomodular lattice is
exactly that one which maps the simplest operations defined on
elements of Hilbert space. Once we have a lattice structure it is
actually completely irrelevant which operation we use and this is
what we have shown in Sec.~\ref{sec:oper}: All 96 Beran expressions
of an orthomodular lattice (quantum logic) can be formulated by
``merged'' conjunction and disjunction in such a way that we can
substitute either classical or any of the five quantum conjunctions
or disjunctions for the ``merged'' ones at will. For example in
$a\cup_3b\>=\>((a'\Cup (b\Cup a))\Cap (((b\Cap a)\Cup (a\Cap b'))\Cup
((a\Cup b)\Cap a')))$ we can substitute either $\cup$ and $\cap$,
or $\cup_1$ and $\cap_1$, or $\cup_5$ and $\cap_5$, etc., for
$\Cup$ and $\Cap$, respectively, and we will always get $a\cup_3b$
on the left hand side. Hence, it is not ``logical properties'' of
operations but lattice structure what is essential. This lattice
structure of Hilbert space we will most probably soon use  as
a basis for the architecture of a would-be quantum computer in a
similar way we use the Boolean algebra for a classical computer.
Here, an equational Hilbert lattice theory partially or even
completely equivalent to the Hilbert space theory would be
an appropriate tool for constructing finite dimensional quantum
logic for quantum computers. Of such equations, one class of
state-determined orthomodular lattice equations of $n$-th order has been
known for 20 years \cite{godow}.  The second such class of Hilbert
operator determined orthomodular lattice
equations---orthoarguesian equations of n-th order---was
found only recently$^{(22,23)}$. Very recently
we found a third such class---possibly the last one.  All these
equations are structurally determined by the properties of
Hilbert space. No semantic considerations whatsoever enter any of the
algorithms which served for finding the equations.

\bigskip\bigskip
\bigskip
\bigskip
\parindent=0pt
{\large\bf ACKNOWLEDGMENT}

\parindent=20pt
\bigskip

One of us (M.~P.) would like to acknowledge support of the Ministry of
Science of Croatia through the Project No.~082006.

\bigskip\bigskip\bigskip
\bibliographystyle{is-abbrv}

\end{document}